%Paper: gr-qc/9508030
%From: ESPOSITO@napoli.infn.it
%Date: Mon, 14 Aug 1995 15:06:30 +0200 (CET-DST)

% File di definizioni GR92.STY

\magnification=1200
\voffset=1.5truecm
\font\tit=cmbx12 scaled\magstep 1
\font\abst=cmsl9

\font\aun=cmbx10
\font\rauth=cmcsc10
\font\subs=cmbxti10
\topskip=1truecm
\hsize=6truein
\vsize=8.5truein
\newcount\who
\who=0

\def\speak{}
\def\centra#1{\vbox{\rightskip=0pt plus1fill\leftskip=0pt plus1fill #1}}

\def\title#1{\baselineskip=20truept\parindent=0pt\centra{\tit #1}
\bigskip\baselineskip=12pt\centra{by}\def\titolo{#1}}

\def\short#1{\def\titolo{#1}}

\def\first#1#2{{\aun #1$^{#2}$}\global\def\firstauthor{#1}}

\def\authors#1{\bigskip\centra{#1}}
\long\def\addresses#1{\bigskip\centra{#1}}
%\vbox{\parindent=15truept\par\leftskip=15truept #1}}
\long\def\addr#1#2{$^{#1}${\it #2}\par}
\long\def\support#1#2{\footnote{}{\hbox to 15truept{\hfill$^{#1}$\ }\sl #2}}

\parskip=5pt
\long\def\summary#1{\bigskip\centra{\speak}\bigskip\medskip
\vbox{\par\leftskip=30truept\rightskip=30truept
\noindent{\bf Summary:} \abst #1}\parindent=15truept}
\def\section#1#2{\who=1\bigskip\medskip
\goodbreak{\bf\noindent\hbox to 15truept{
   #1.\hfil} #2}\nobreak
\medskip\nobreak\who=0}
\def\subsection#1#2{\ifnum\who=0\bigskip\goodbreak\else\smallskip\fi
{\subs\noindent\hbox to 25truept{#1.\hfil} #2}\nobreak
\ifnum\who=0\medskip\fi\nobreak}
\def\references{\bigskip\medskip
\goodbreak{\bf\noindent\hskip25truept References
   }\nobreak
\medskip\nobreak
\frenchspacing\pretolerance=2000\parindent=25truept}
\def\paper#1#2#3#4#5#6{\item{\hbox to 20truept{[#1]\hfill}}
{\rauth #2,} {\it #3} {\bf #4} (#5) #6\smallskip}
\def\book#1#2#3#4#5{\item{\hbox to 20truept{[#1]\hfill}} {\rauth #2,} {\it #3},
   #4 (#5)\smallskip}

\newcount\firstp
\firstp=\pageno

\headline={\ifnum\pageno=\firstp\hfill\else
\line{\firstauthor,\quad{\it \titolo}\hfill\folio}\fi}
\footline={\line{\hfill}}

\title{THE GEOMETRY OF COMPLEX SPACE-TIMES WITH TORSION}

\short{The Geometry of Complex Space-Times with Torsion}

\authors{
\first{Giampiero ESPOSITO}{1,2}}

\addresses{
\addr{1}{International Centre for Theoretical Physics,
Strada Costiera 11, 34014 Trieste, Italy}
\addr{2}{Scuola Internazionale Superiore di Studi Avanzati,
Via Beirut 2-4, 34013 Trieste, Italy}}

\speak{}

\summary{The necessary and sufficient condition for the
existence of $\alpha$-surfaces in complex space-time manifolds
with nonvanishing torsion is derived. For these manifolds,
Lie brackets of vector fields and spinor Ricci
identities contain explicitly the effects of torsion. This
leads to an integrability condition for
$\alpha$-surfaces which does not
involve just the self-dual Weyl spinor, as in
complexified general relativity, but
also the torsion spinor, in a nonlinear way, and its covariant
derivative.
A similar result also holds for four-dimensional,
smooth real manifolds with a positive-definite metric.
Interestingly, a particular solution of the integrability condition is given
by right-flat and right-torsion-free space-times.}

Twistor theory was created by Roger Penrose as an
approach to the quantum-gravity problem where null surfaces
and some particular complex manifolds are regarded as
fundamental entities, whereas space-time points are only derived
objects, since they might become ill-defined at the Planck length
[1,2]. It is by now well-known that the building blocks of classical
field theory in Minkowski space-time are the $\alpha$-planes. In other
words, one first takes complexified compactified Minkowski space
$CM^{\#}$, and one then defines $\alpha$-planes as null two-surfaces,
such that the metric vanishes over them, and their null tangent vectors
have the two-component-spinor form $\lambda^{A}\pi^{A'}$, where
$\lambda^{A}$ is varying and $\pi^{A'}$ is fixed by a well-known
differential equation. This definition can be generalized to complex
or real Riemannian (i.e. with positive-definite metric) space-times
provided the Weyl curvature is anti-self-dual, giving rise to the
so-called $\alpha$-surfaces.

We are here interested in the case of complex space-times with
nonvanishing torsion (hereafter referred to as $CU_{4}$ space-times).
This study appears relevant at least for the following reasons:
(i) torsion is a peculiarity of relativistic theories
of gravitation; (ii) the gauge theory of the Poincar\'e group leads
to theories with torsion; (iii) theories with torsion are theories of
gravity with second-class constraints; (iv) if torsion is nonvanishing,
the occurrence of cosmological singularities can be less generic than in
general relativity [3]. We have thus studied the problem: what are the
conditions on curvature and torsion for a $CU_{4}$ space-time to
admit $\alpha$-surfaces (defined as above) ?

The starting point of our calculation, based on the use of the full
$U_{4}$-connection with torsion [2], is the evaluation of the Lie
bracket of two vector fields $X$ and $Y$ tangent to a totally null
two-surface in $CU_{4}$. By virtue of Frobenius' theorem, this Lie bracket
is a linear combination of $X$ and $Y$:
$$
[X,Y]=\varphi X + \rho Y
\; \; \; \; ,
\eqno (1)
$$
where $\varphi$ and $\rho$ are scalar functions. Moreover, using the
definition of the torsion tensor $S(X,Y)$, one also has:
$$
[X,Y] \equiv \nabla_{X}Y -\nabla_{Y}X -2S(X,Y)
\; \; \; \; .
\eqno (2)
$$
Thus, since in $CU_{4}$ models the torsion tensor,
antisymmetric in the first two indices, can be expressed spinorially
as:
$$
S_{ab}^{\; \; \; c}=\chi_{AB}^{\; \; \; \; \; CC'}\epsilon_{A'B'}+
{\widetilde \chi}_{A'B'}^{\; \; \; \; \; \; \; CC'}
\epsilon_{AB}
\; \; \; \; ,
\eqno (3)
$$
where the spinors $\chi$ and ${\widetilde \chi}$ are symmetric
in $AB$ and $A'B'$ respectively, and are {\it totally independent},
one finds by comparison of Eqs. (1) and (2) that [2]:
$$
\pi^{A'}\Bigr(\nabla_{AA'}\pi_{B'}\Bigr)=\xi_{A}\pi_{B'}
-2\pi^{A'}\pi^{C'}{\widetilde \chi}_{A'B'AC'}
\; \; \; \; ,
\eqno (4)
$$
where we have set $X^{a}=\lambda^{A}\pi^{A'}$, $Y^{a}=\mu^{A}\pi^{A'}$.
This is the desired necessary and sufficient condition for the field
$\pi^{A'}$ to define an $\alpha$-surface in the presence of torsion [1,2].
We now have to derive the integrability condition for Eq. (4). For this
purpose, we operate with $\pi^{B'}\pi^{C'}\nabla_{\; \; C'}^{A}$ on
both sides of Eq. (4), we repeatedly use
the Leibniz rule and Eq. (4), and we
also use the spinor formula for the Riemann tensor and spinor Ricci
identities. The relations we need are respectively [2]:
$$ \eqalignno{
R_{abcd}&=\psi_{ABCD}\epsilon_{A'B'}\epsilon_{C'D'}
+{\widetilde \psi}_{A'B'C'D'}\epsilon_{AB}\epsilon_{CD}\cr
&+\Phi_{ABC'D'}\epsilon_{A'B'}\epsilon_{CD}
+{\widetilde \Phi}_{A'B'CD}\epsilon_{AB}\epsilon_{C'D'}\cr
&+\Sigma_{AB}\epsilon_{A'B'}\epsilon_{CD}\epsilon_{C'D'}
+{\widetilde \Sigma}_{A'B'}\epsilon_{AB}\epsilon_{CD}
\epsilon_{C'D'}\cr
&+\Lambda \Bigr(\epsilon_{AC}\epsilon_{BD}+\epsilon_{AD}
\epsilon_{BC}\Bigr)\epsilon_{A'B'}\epsilon_{C'D'}\cr
&+{\widetilde \Lambda}\Bigr(\epsilon_{A'C'}
\epsilon_{B'D'}+\epsilon_{A'D'}\epsilon_{B'C'}\Bigr)
\epsilon_{AB}\epsilon_{CD}
\; \; \; \; ,
&(5)\cr}
$$
$$
\left[\nabla_{C(A'}\nabla_{\; \; B')}^{C}
-2{\widetilde \chi}_{A'B'}^{\; \; \; \; \; \; \; HH'}
\nabla_{HH'}\right]\pi^{C'}
={\widetilde \psi}_{A'B'E'}^{\; \; \; \; \; \; \; \; \; \; \; C'}
\; \pi^{E'}
-2{\widetilde \Lambda}\pi_{(A'}\epsilon_{B')}^{\; \; \; \; C'}
+{\widetilde \Sigma}_{A'B'}\pi^{C'}
\; \; \; \; .
\eqno (6)
$$
As usual, the {\it twiddle} symbol denotes spinor or scalar quantities
independent of their untwiddled counterpart. The spinors $\psi$ and
${\widetilde \psi}$ are the Weyl spinors, and are thus invariant under
conformal rescalings of the metric [1]. The spinors $\Sigma$ and
${\widetilde \Sigma}$ express the antisymmetric part of the Ricci
tensor. Using also the well-known property $\lambda_{A}\lambda^{A}=
\pi_{A'}\pi^{A'}=0$, the integrability condition for $\alpha$-surfaces
is thus found to be [2]:
$$ \eqalignno{
{\widetilde \psi}_{A'B'C'D'}&=
-4{\widetilde \chi}_{A'B'AL'}
{\widetilde \chi}_{C' \; \; \; \; \; \; D'}^{\; \; \; \; L'A}
+4{\widetilde \chi}_{L'B'AC'}
{\widetilde \chi}_{A'D'}^{\; \; \; \; \; \; \; \; AL'}\cr
&+2\nabla_{\; \; D'}^{A}\Bigr({\widetilde \chi}_{A'B'AC'}\Bigr)
\; \; \; \; .
&(7)\cr}
$$
Note that an analogous result holds if the metric is positive-definite
rather than complex [2]. A four-manifold with nonvanishing torsion
and positive-definite metric is here denoted by $RU_{4}$.
Interestingly, a particular solution of Eq. (7) is given
by ${\widetilde \psi}_{A'B'C'D'}=0$, ${\widetilde \chi}_{A'C'AB'}=0$.
By analogy with complexified general relativity, the particular
$CU_{4}$ and $RU_{4}$ space-times satisfying these additional conditions
are here called right-flat and right-torsion-free. This means that the
surviving Weyl and torsion spinors, i.e. $\psi_{ABCD}$ and
$\chi_{AB}^{\; \; \; \; \; CC'}$, do not affect the integrability
condition (7) for $\alpha$-surfaces.

\bigskip\medskip
{\bf\noindent\hskip 25truept Acknowledgments}
\medskip

I am very grateful to the referees at GRG for correcting a large
number of errors affecting the original version of a paper of
mine, which I later revised in a substantial way, writing Ref. [2].
The successful completion of my research, motivated by Ref. [2],
would have been unconceivable
without the substantial help I received from these two anonymous
referees. I am also very grateful to Professor Abdus Salam,
the International Atomic Energy Agency and UNESCO for hospitality
at the ICTP, and to Professor
Dennis Sciama for hospitality at the SISSA of Trieste. Last but
not least, I thank Paul Haines for finding a misprint in
the manuscript.

\references

\book{1}{R. S. Ward, R. O. Wells}{Twistor Geometry and Field
Theory}{Cambridge Monographs on Mathematical Physics.}
{Cambridge University Press, Cambridge, 1990}

\paper{2}{G. Esposito}{DSF preprint}{n. 20}{1991}
{INFN and Dipartimento di Scienze Fisiche di Napoli.}

\paper{3}{G. Esposito}{Fortschr. der Physik}{40}{1992}{1-30.}
\bye